\documentclass[%        Class options:
aps,%                   American Physical Society
prd,%                   Physical Review 
tightenlines,%          Single spaced lines
superscriptaddress,%    Authors' addresses linked with superscripts
nofootinbib,%           Does not treat footnotes as references
floatfix]%              Fixes float errors
{revtex4}%              REVTEX 4 Package used
\usepackage{graphicx,%  Default Latex 2eps package for embedding figures
longtable,%              Useful for long table
color}

\begin{document}
\title{Evidence of $\theta_{13}>0$ from global neutrino data analysis}
\author{		G.L.~Fogli}
\affiliation{   	Dipartimento Interateneo di Fisica ``Michelangelo Merlin,'' %\\
               		Via Amendola 173, 70126 Bari, Italy}
\affiliation{   	Istituto Nazionale di Fisica Nucleare, Sezione di Bari, %\\
               		Via Orabona 4, 70126 Bari, Italy}
\author{		E.~Lisi}
\affiliation{   	Istituto Nazionale di Fisica Nucleare, Sezione di Bari, %\\
               		Via Orabona 4, 70126 Bari, Italy}
\author{		A.~Marrone}
\affiliation{   	Dipartimento Interateneo di Fisica ``Michelangelo Merlin,'' %\\
               		Via Amendola 173, 70126 Bari, Italy}
\affiliation{   	Istituto Nazionale di Fisica Nucleare, Sezione di Bari, %\\
               		Via Orabona 4, 70126 Bari, Italy}
\author{		A.~Palazzo}
\affiliation{ 		Cluster of Excellence, Origin and Structure of the Universe, 
					  Technische Universit\"at M\"unchen,
					  Boltzmannstra\ss{e} 2, D-85748, Garching, Germany}
\author{		A.M.~Rotunno}
\affiliation{   	Dipartimento Interateneo di Fisica ``Michelangelo Merlin,'' %\\
               		Via Amendola 173, 70126 Bari, Italy}
%\date{{\today}}

\begin{abstract}
The neutrino mixing angle $\theta_{13}$ is at the focus of current neutrino research. 
From a global analysis of the available oscillation data in a $3\nu$ framework, we previously reported 
{[Phys.\ Rev.\ Lett.\ {101}, 141801 (2008)]}
hints in favor of $\theta_{13}>0$ 
at the 90\% C.L.  Such hints are consistent with 
the recent indications of $\nu_\mu\to\nu_e$ appearance in the T2K and MINOS
long-baseline accelerator experiments. Our global analysis of all the available data 
currently provides $>3\sigma$ evidence for nonzero $\theta_{13}$, with $1\sigma$ ranges   
$\sin^2\theta_{13}=0.021\pm0.007$ or $0.025\pm0.007$, depending on reactor neutrino flux systematics.
Updated ranges are also reported for the other $3\nu$ 
oscillation parameters $(\delta m^2,\,\sin^2\theta_{12})$ and $(\Delta m^2,\,\sin^2\theta_{23})$.
\end{abstract}
\maketitle

%%%%%%%%%%%%%%%%%%%%%%%%%%%%%%%%%%%%%%%%%%%%%%%%%%%%%%%%%%%%%%%%%%%%%%
%%%% Section I %%%%%%%%%%%%%%%%%%%%%%%%%%%%%%%%%%%%%%%%%%%%%%%%%%%%%%%
%%%%%%%%%%%%%%%%%%%%%%%%%%%%%%%%%%%%%%%%%%%%%%%%%%%%%%%%%%%%%%%%%%%%%%

\section{Introduction}

Neutrino oscillation experiments have established that neutrino flavor and mass states do mix \cite{Na10}.
In the simplest framework, the three flavor states  $(\nu_e,\nu_\mu,\nu_\tau)$ 
are quantum superpositions of only three light mass states $(\nu_1,\,\nu_3,\,\nu_3)$ 
via a unitary mixing matrix $U$, parametrized in terms
of three rotation angles $(\theta_{12},\theta_{13},\theta_{23})$
and one possible CP-violating phase $\delta$ in standard notation \cite{Na10}. 
The amplitudes and frequencies of 
flavor oscillation phenomena are governed, respectively, by the $\theta_{ij}$
angles and by two squared mass differences, namely,
\begin{equation}
\label{dm2}
\delta m^2=m^2_2-m^2_1>0
\end{equation}
and, in our convention \cite{Fo06}, 
\begin{equation}
\label{Dm2}
\Delta m^2=m^2_3-\frac{m^2_2+m^2_1}{2}\ , 
\end{equation}
where $\Delta m^2>0$ ($<0$) corresponds to normal (inverted) mass spectrum hierarchy. Typically, a single
experiment is mainly sensitive to only one of the above mass gaps and to one mixing parameter, although
subleading effects driven by the remaining parameters may become relevant in precision oscillation searches.

So far, solar and long-baseline reactor neutrino experiments have measured the mass-mixing parameters ($\delta m^2,\,\theta_{12}$) in the
$\nu_e\to\nu_e$ channel, while atmospheric 
and long-baseline accelerator (LBL) experiments have measured 
($\Delta m^2,\,\theta_{23}$) in the $\nu_\mu\to\nu_\mu$ channel.
Conversely, short-baseline reactor experiments, 
mainly sensitive to $(\Delta m^2,\,\theta_{13})$, have 
set upper---but not lower---bounds on the mixing angle $\theta_{13}$; see \cite{Na10} for an overview.
However, we observed in \cite{NOVE,HINT} that  the two data sets mainly sensitive to $\delta m^2$ 
and to $\Delta m^2$ provided  two separate hints in favor of  
$\theta_{13}>0$ which, in combination, disfavored the null hypothesis $\theta_{13}=0$ at 90\% C.L.\ \cite{HINT}.

The statistical significance of the hints, 
as well as their possible origin in subleading oscillation effects,
have been examined in detail and also debated in a number of analyses 
\cite{Baha,Adde,Sc08,Baye,Crit,Lati,Ve09,SNOL,Go10,We10,Me10,KL11,Nu10}, often triggered 
by new input data and, more recently, also by a new, critical evaluation of older inputs
for the reactor neutrino fluxes \cite{Ano1} (see \cite{Ano2,Hube,Sc11,Ve11}). 
Within the standard $3\nu$ framework (with no extra sterile neutrinos $\nu_s$), 
the overall statistical significance of $\theta_{13}>0$ has, so far, not exceeded 
the level of $\sim 2\sigma$, with a corresponding estimated range 
$\sin^2\theta_{13}\simeq 0.02\pm0.01$, see e.g.\ \cite{Ve09,Ve11}.  

Very recently (June 2011),
new relevant results have been announced by  two  
long-baseline accelerator experiments probing the $\nu_\mu\to\nu_e$ appearance channel, 
which is governed
by the $(\Delta m^2,\,\theta_{13})$ parameters (although with an
additional dependence on $\theta_{23}$ and $\delta$, absent
in short-baseline reactor experiments). In particular, the 
Tokai-to-Kamioka (T2K) experiment has observed 6 electron-like events with
an estimated background of 1.5 events, rejecting 
$\theta_{13}=0$ at the level of $2.5\sigma$ \cite{TtoK}. The low background  level makes
the T2K results particularly important and robust.
Shortly after, 
the Main Injector Neutrino Oscillation Search (MINOS) experiment has reported 
the observation of 62 electron-like events with
an estimated background of 49 events, disfavoring
$\theta_{13}=0$ at $1.5\sigma$ \cite{MINO,MIN2}.
Taken together, these data suggest $\sin^2\theta_{13}\simeq \mathrm{few}\;\%$,
in agreement with our previous hints \cite{HINT} discussed above.

It makes then sense to update our previous global analyses of oscillation data \cite{HINT,Adde,Ve09} by 
including the latest T2K and MINOS results, as well as other data which have been
published in the last few years. Our main result is
%...................................
\begin{equation}
\sin^2\theta_{13}=\left\{ 
\begin{array}{l}
0.021\pm 0.007\, ,\ \mathrm{old\ reactor\ fluxes}\\[1mm]
0.025\pm 0.007\, ,\ \mathrm{new\ reactor\ fluxes}
\end{array}
\right. \ (1\sigma)\ , 
\end{equation}
%....................................... 
corresponding to a $>3\sigma$ evidence in favor of nonzero $\theta_{13}$ 
(while previous hints did not exceed the $\sim 2\sigma$ level). We discuss below  some details of 
our current analysis, and a few relevant implications of $\theta_{13}>0$ for neutrino physics.

\section{Input data and their analysis}

We briefly summarize the main updates in our global analysis of neutrino oscillations
driven by $\delta m^2$ and by $\Delta m^2$, respectively, with respect to our previous
works on the subject.

\subsection{Data sensitive to $\delta m^2$}

These data include events collected in solar neutrino experiments and in the long-baseline reactor experiment
KamLAND. As described in \cite{Adde,GEON},  we take as free parameters
$(\delta m^2,\,\theta_{12},\,\theta_{13})$, as well as the four geoneutrino event rates
corresponding to decays of natural Thorium and Uranium in the KamLAND and Borexino experiments, which
are then marginalized away. With respect to \cite{GEON}, we update the KamLAND results on reactor
and geoneutrino data \cite{KL11,In10}, and include
the Borexino {\color{black} total event rates of } $^7$Be \cite{BoBe} and $^8$B \cite{BoBi} solar neutrinos.
As a side result, we obtain good agreement with the latest geoneutrino analysis performed
by the KamLAND collaboration \cite{In10}. 

The interpretation of both solar and reactor neutrino data depend, to some extent, on the estimated fluxes (and their
uncertainties) at their sources, namely, the solar and reactor cores, respectively. 
Concerning solar neutrinos, the long-standing issue of high versus low metallicity (see \cite{Meta} for
a review of the subject) affects only marginally the estimates of the oscillation parameters 
{\color{black} \cite{Sc08,Go10}. For instance, we find that by using the standard solar model (SSM) inputs 
GS98 and AGSS09 from Table~3 of \cite{Meta}, the significance of 
the solar+KamLAND constraints on $\sin^2\theta_{13}$ changes by only 
$\sim 0.1\sigma$. For definiteness, we use the reference BP05(OP) SSM from Table~2 of \cite{Bahc}, which
gives fit results intermediate between the GS98 or AGSS09 SSM.} 

Conversely, a more relevant issue has been recently pointed out in the context of
reactor neutrino fluxes, whose detailed reevaluation in \cite{Ano1} 
suggests an average increase of about $3.5\%$ in  normalization, with respect to previously accepted
standards \cite{Grat} {\color{black} (see also \cite{Hube})}. The increase would then
indicate extra electron flavor disappearance  in KamLAND \cite{Ano2}
which, in the context of $3\nu$ oscillations, are expected
to induce a small but nonnegligible shift in $\theta_{12}$ and $\theta_{13}$ \cite{Ve11}. 
{\color{black} We will thus present
results obtained by using either ``old'' \cite{Grat} and ``new'' \cite{Ano1} reactor fluxes and
their uncertainties in the analysis of reactor data, in order to illustrate the 
effect of such flux systematics.}

\subsection{Data sensitive to $\Delta m^2$}

These data include, in our analysis, events collected in the atmospheric neutrino experiment Super-Kamiokande (SK),
in the short-baseline reactor experiment CHOOZ, and in the long-baseline experiments K2K, T2K, and MINOS.

With respect to our previous analysis \cite{Fo06}, the CHOOZ data \cite{CHOO} are analyzed
also with new input reactor fluxes \cite{Ano1}, whose higher normalization leads to a small disappearance
effects which slightly favors nonzero $\theta_{13}$ and weakens its upper bounds \cite{Ano2}.
 
We update the atmospheric neutrino analysis \cite{Fo06,Adde} by including SK-II and SK-III data \cite{We10,Nu10}, 
together with
improved estimates for the associated systematic errors \cite{Thes}. However, a reduction of the SK-I+II+III data is unavoidable,
since the official SK analysis includes many categories of events, bins, and systematics, which cannot be fully
reproduced outside the collaboration, as already noticed \cite{Ve11}. 
In particular, we exclude from
our analysis multi-ring and pion-decay events, and we group together some classes of partially contained events
(stopping and through-going) and of upgoing muon events
(showering and nonshowering).
 We think that, despite some unavoidable approximations and 
loss of information, our analysis remains a useful,
independent study of SK atmospheric data within a full $3\nu$ oscillation framework. A similar
attitude has been adopted in \cite{Go10}, while the authors in \cite{Sc11} have directly taken the official
SK likelihood function for the $(\Delta m^2,\,\theta_{23},\,\theta_{13})$ parameters
from  a $3\nu$ analysis with $\delta m^2$ set to zero \cite{We10}. As a consequence, the atmospheric
neutrino likelihood adopted in \cite{Sc11} lacks the $3\nu$ subleading oscillation effects 
driven by $\delta m^2$ \cite{Pere}, which are potentially relevant at the current level of
accuracy \cite{Fo06,Nu10}.

Concerning long-baseline accelerator data in the $\nu_\mu\to\nu_\mu$ disappearance channel, 
we maintain our previous K2K analysis without changes \cite{Fo06,Adde}, but we update the MINOS spectrum
analysis from \cite{Disa}. We do not include MINOS antineutrino disappearance data, 
which seem to indicate a puzzling deviation from the neutrino
best-fit parameters ($\Delta m^2,\,\theta_{23}$) \cite{Anti}. The deviation should
be carefully monitored in the future, since it cannot be interpreted in the standard $3\nu$ framework.

In the appearance channel $\nu_\mu\to\nu_e$, we include the recent results from \cite{MINO,MIN2}, by
fitting the total (spectrum-integrated) rate of electron-like events at the far detector, 
taking into account its statistical and systematic
errors. We reproduce with good accuracy the marginalized MINOS confidence level contours \cite{MIN2}
in the plane charted by 
the CP phase $\delta$ and by the dominant mixing parameter $\sin^2\theta_{23}\sin^22\theta_{13}$,
for both normal and inverted hierarchy. 
Last, but definitely not least, the crucial T2K appearance data recently reported in \cite{TtoK} 
are included by fitting the 
total (spectrum integrated) rate at the far detector within Poisson statistics. 
We reproduce with good accuracy the T2K confidence level contours in the plane charted by 
the CP phase $\delta$ and by the mixing parameter $\sin^22\theta_{13}$ for fixed values 
of the other oscillation parameters, in both normal and inverted hierarchy \cite{TtoK}. 
{\color{black} [Note that all such fixings are removed in the global analysis below.]

We emphasize that the official appearance data analyses in T2K \cite{TtoK} and MINOS
\cite{MIN2} include also spectral information and uncertainties, which are
quite difficult to model outside the respective collaborations. However, we think that
their current parameter fits can be reproduced, to a large extent, by
a single number: the total appearance rate (normalized to published no-oscillation expectations in
each experiment). 
In particular, we have not found appreciable variations by performing trial 
fits with binned spectra.
We also note that the continuous degeneracy of the $(\theta_{13},\,\delta)$ minima, 
trivially expected in a total-rate fit, is not lifted by the additional spectral information of
complete fits \cite{TtoK,MIN2}. Of course, future, higher-statistics 
T2K and MINOS results might entail more constraining spectral information than in current data.}

\vspace*{-.3cm}
\subsection{Three-neutrino analysis}

In $\delta m^2$-sensitive oscillation searches, the relevant $3\nu$ 
variables are $(\delta m^2,\,\sin^2\theta_{12},\,\sin^2\theta_{13})$.
A very minor dependence of solar neutrinos on $\pm \Delta m^2$ \cite{Hier,Fo06} is 
also kept for the sake of precision.

In $\Delta m^2$-sensitive oscillation searches, we take as free parameters $(\pm\Delta m^2,\,\sin^2\theta_{23},\,
\sin^2\theta_{13})$, while  ($\delta m^2,\,\theta_{12}$) are fixed at their best-fit values from
the analysis of $\delta m^2$-sensitive data. Our atmospheric $\nu$ analysis is  restricted to the two CP-conserving
cases $\cos\delta=\pm 1$ \cite{Fo06,Adde}; we plan to remove such restriction in the future, so as to analyze both
atmospheric and long-baseline accelerator data with unconstrained $\delta$ 
{\color{black} (see \cite{Go10} for a previous
analysis of this kind). Anyway, as argued at the end of Sec.~III, the evidence
of $\theta_{13}>0$ can only be strengthened for unconstrained $\delta$.}

In the global fit, note that the common parameter $\sin^2\theta_{13}$ is constrained by both classes of
oscillation searches. As in previous works \cite{Fo06,Adde,Ve09}, we will present bounds on the continuous parameters 
\begin{equation}
\label{cont}
(\delta m^2,\,\sin^2\theta_{12},\,\sin^2\theta_{13},\,\sin^2\theta_{23},\,\Delta m^2)\ ,
\end{equation}
marginalized over the four discrete cases 
\begin{equation}
\label{disc}
[\mathrm{sign}(\Delta m^2)=\pm1]\otimes [\cos\delta=\pm 1]\ ,
\end{equation} 
unless otherwise stated. Concerning $\Delta m^2$, note that our convention in Eq.~(\ref{Dm2}) 
absorbs the trivial $\delta m^2$ difference arising in the best-fit 
values (for normal and inverted hierarchy) of the alternative conventional mass parameters 
$\Delta m^2_{31}=m^2_3-m^2_1$ or $\Delta m^2_{32}=m^2_3-m^2_2$. 

A final remark is in order, concerning some issues recently raised by a reevaluation of reactor fluxes \cite{Ano1,Ano2}. 
The higher normalization suggested in \cite{Ano1}
produces extra disappearance in KamLAND and CHOOZ, which jointly favor higher $\theta_{13}$ and $\theta_{12}$
in a $3\nu$ framework. However, it also produces a small disappearance effect in very short-baseline
experiments (not included in our fit) which is at variance with standard $3\nu$ oscillations, and suggests
hypothetical sterile neutrino oscillation driven by a mass gap $\Delta M^2\sim O(1)$~eV$^2$ and by relatively small 
 mixing with $\nu_e$ \cite{Ano2,Ko11,Pa11,Ro11,Gi11}.
Conversely, the old normalization is essentially consistent with the $3\nu$ scenario, but slightly weakens
the global indications for $\theta_{13}>0$. Our educated guess (or prejudice) is that the true normalization may be 
intermediate between the old and the new one, and that previous normalization errors were 
 underestimated. Anyway,
we shall show  results for both old and new reactor fluxes, 
their difference being an indication of the corresponding systematic uncertainties. 
{\color{black} We emphasize that the understanding of these puzzling systematics 
definitely requires further experimental data. We also note that fluxes even lower than the ``old'' ones 
have been investigated as a possible phenomenological
explanation of very short baseline reactor data \cite{Ano2,Sc11}. In general,
the lower the reactor fluxes, the lower the allowed values of $\theta_{13}$, as discussed below.}

\vspace*{-.3cm}
\section{Results}

Figure~1 shows the main results of our global fit, in terms of allowed ranges 
for each of the oscillation parameters in Eq.~(\ref{cont}), as obtained by marginalizing 
the remaining parameters over all the four discrete cases in Eq.~(\ref{disc}). The vertical scale
represents the number of standard deviations ($N\sigma=\sqrt{\Delta\chi^2}$) from the best fit point. The lines departing from each best fit
would be perfectly straight and symmetric for exactly gaussian error distributions.
This is nearly the case for all but the $\sin^2\theta_{23}$ parameter, which shows a skewed preference
for the first octant at the level of $1\sigma$, in qualitative agreement with our 
previous analyses \cite{Fo06,Adde}. The estimates of $\sin^2\theta_{13}$ and
$\sin^2\theta_{12}$ are affected by reactor flux systematics; in particular, the  
dashed lines refer to the analysis with new reactor fluxes \cite{Ano1}, which turn out to 
shift both parameters by roughly $+0.005$ with respect to the
case with old reactor fluxes (solid lines). In both cases, however, there is  
a clear evidence in favor of $\sin^2\theta_{13}>0$, at a confidence level of at 
least $3\sigma$. This evidence is the  most important outcome of our work.

\newpage

%%%%%%%%%%%%%%%%%%%%%%%%%%%%%%%%%%%%%%%%%%%%%%%%%%%%%%%%%%%%%%%%%%%%%
\begin{figure}[t]
\vspace*{-1.0cm}
\hspace*{-1.0cm}
\includegraphics[width=1.1\textwidth]{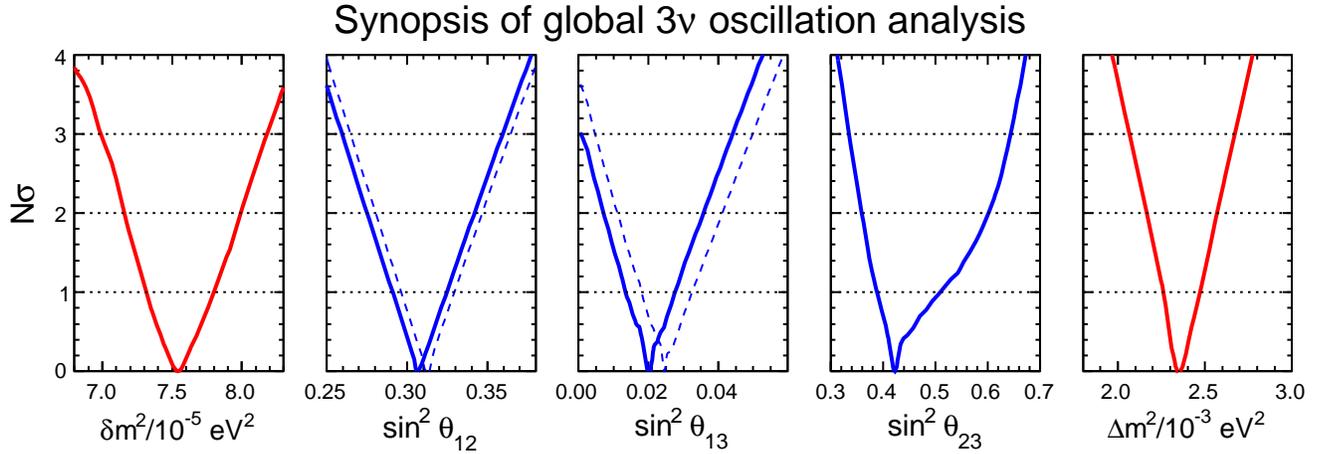}
\vspace*{-2.0cm}
\caption{\label{fig_1} 
Global $3\nu$ oscillation analysis. Bounds on the 
mass-mixing oscillation parameters, in terms of standard deviations  from 
the best fit. Solid (dashed) lines refer to the case with
old (new) reactor neutrino fluxes. Note the 
$>3\sigma$ evidence for $\theta_{13}>0$.}
\end{figure}
%%%%%%%%%%%%%%%%%%%%%%%%%%%%%%%%%%%%%%%%%%%%%%%%%%%%%%%%%%%%%%%%%%%%%%%

Figure~2 breaks down the global evidence for $\sin^2\theta_{13}>0$ into two separate contributions coming
from the data sets sensitive to either $\delta m^2$ (Solar+KamLAND) or $\Delta m^2$ (ATM+LBL+CHOOZ),
assuming old and new reactor fluxes (left and right panels, respectively). Remarkably, the two
data sets agree very well, with best fits rather close to each other in both panels,
and with nearly gaussian uncertainties in all cases.
The bounds from combined (ALL) data appear to be currently dominated by $\Delta m^2$-sensitive experiments---not surprisingly, since the T2K appearance results alone account for more than $2\sigma$ \cite{TtoK}. 
The T2K experiment, currently limited by  
statistics rather than by systematics, is expected to improve significantly the
bounds on $\theta_{13}$ in future physics runs \cite{TtoK}. 
We also find it useful to summarize the $\pm 1\sigma$ ranges of $\sin^2\theta_{13}$ 
in a different format in Fig.~3, where the solid and dashed
error bars refer to old and new reactor neutrino fluxes, respectively.

%%%%%%%%%%%%%%%%%%%%%%%%%%%%%%%%%%%%%%%%%%%%%%%%%%%%%%%%%%%%%%%%%%%%%
\begin{figure}[b]
\vspace*{-.6cm}
\includegraphics[width=0.99\textwidth]{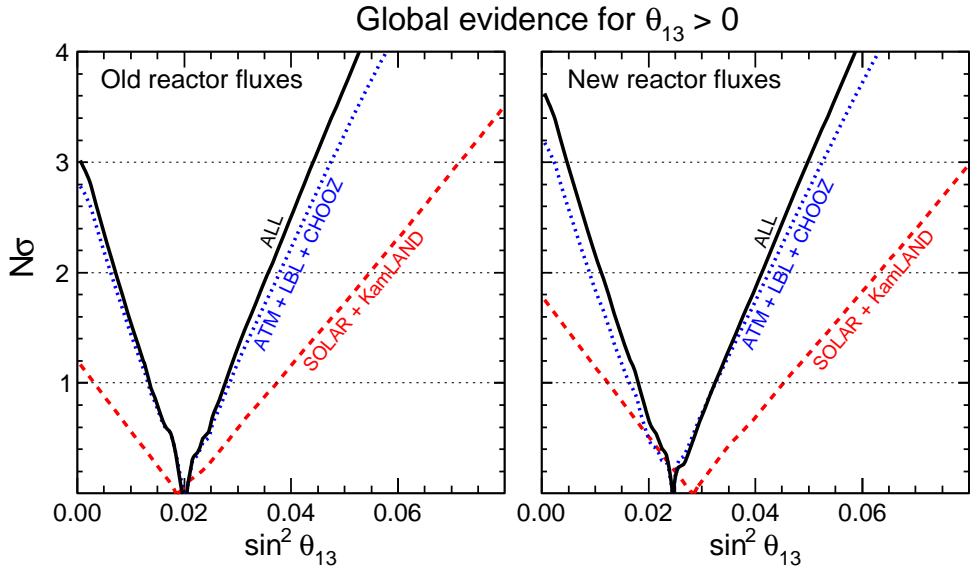}
\vspace*{-1.cm}
\caption{\label{fig_2} 
Breakdown of the evidence for $\theta_{13}>0$ from the global fit (ALL) 
into contributions coming from $\delta m^2$-sensitive data (Solar+KamLAND)
and from $\Delta m^2$-sensitive data (ATM+LBL+CHOOZ). The left and right panels
refer to old and new fluxes, respectively.}
\end{figure}
%%%%%%%%%%%%%%%%%%%%%%%%%%%%%%%%%%%%%%%%%%%%%%%%%%%%%%%%%%%%%%%%%%%%%%%

\newpage

%%%%%%%%%%%%%%%%%%%%%%%%%%%%%%%%%%%%%%%%%%%%%%%%%%%%%%%%%%%%%%%%%%%%%
\begin{figure}[t]
\vspace*{0.cm}
\hspace*{-1.9cm}
\includegraphics[width=1.0\textwidth]{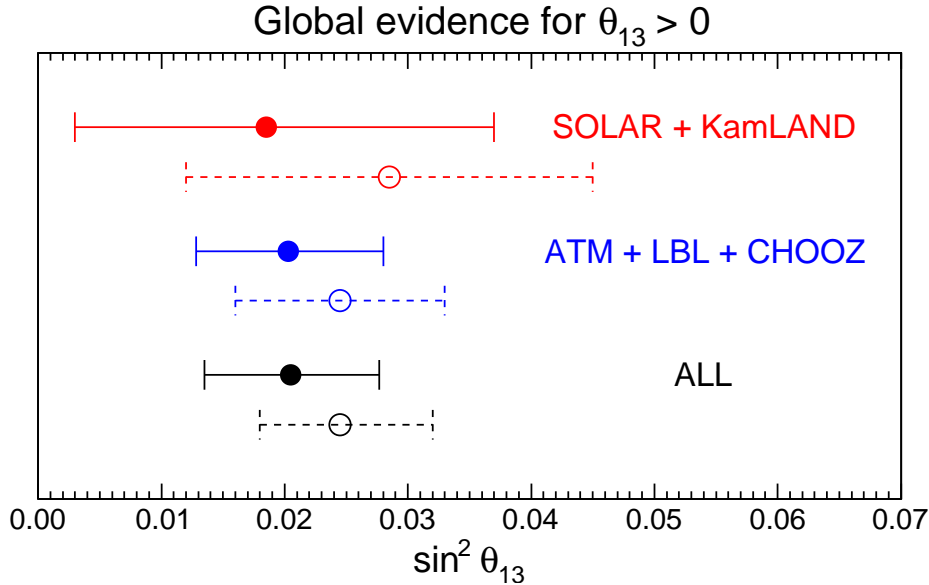}
\vspace*{-.6cm}
\caption{\label{fig_3} 
Global $3\nu$ analysis. Preferred $\pm1\sigma$ ranges for the mixing parameter $\sin^2\theta_{13}$
from partial and global data sets. Solid and dashed error bars refer  
to old and new reactor neutrino fluxes, respectively.}
\end{figure}
%%%%%%%%%%%%%%%%%%%%%%%%%%%%%%%%%%%%%%%%%%%%%%%%%%%%%%%%%%%%%%%%%%%%%%%

%===========================================================================
\begin{table}[b]
\caption{\label{Synopsis} Results of the global $3\nu$ oscillation analysis, in terms of best-fit values and
allowed 1, 2 and $3\sigma$ ranges  for the mass-mixing parameters, assuming old reactor neutrino 
fluxes. By using new reactor fluxes, the corresponding best fits and ranges for $\sin^2\theta_{12}$
and $\sin^2\theta_{13}$ (in parentheses) are basically shifted by about $+0.006$ and $+0.004$, respectively,
while the other parameters are essentially unchanged. All the results are marginalized over the four discrete cases in 
Eq.~(\protect\ref{disc}).}
%\centering
\resizebox{\textwidth}{!}{
\begin{ruledtabular}
\begin{tabular}{cccccc}
Parameter & $\delta m^2/10^{-5}\mathrm{\ eV}^2$ & $\sin^2\theta_{12}$ & $\sin^2\theta_{13}$ & $\sin^2\theta_{23}$ &
$\Delta m^2/10^{-3}\mathrm{\ eV}^2$ \\[4pt]
\hline%---------------------------------------------------------------------
Best fit        &     7.58     &  0.306          &  0.021          &  0.42           &  2.35         \\
                &              & (0.312)         &  (0.025)        &                 &               \\
$1\sigma$ range & 7.32~--~7.80 & 0.291~--~0.324  &  0.013~--~0.028 &  0.39~--~0.50 & 2.26~--~2.47  \\
                &              &(0.296~--~0.329) & (0.018~--~0.032)&                 &               \\
$2\sigma$ range & 7.16~--~7.99 & 0.275~--~0.342  &  0.008~--~0.036 &  0.36~--~0.60 & 2.17~--~2.57  \\
                &              &(0.280~--~0.347) & (0.012~--~0.041)&                 &               \\
$3\sigma$ range & 6.99~--~8.18 & 0.259~--~0.359  &  0.001~--~0.044 &  0.34~--~0.64 & 2.06~--~2.67  \\ 
                &              &(0.265~--~0.364) & (0.005~--~0.050)&                 &
\end{tabular}
\end{ruledtabular}
}%end of resizebox
%\vspace*{.2cm}
\end{table}
%============================================================================

Table~I reports  the bounds shown in Figs.~1--3 in numerical form. All the bounds are largely
uncorrelated from each other; e.g., the allowed ranges of  $\delta m^2$ and $\Delta m^2$ are
basically independent on variations of the mixing angles within their uncertainties (not shown). 
Nevertheless, we find it useful to report the joint ranges for the mixing parameters $\sin^2\theta_{ij}$,
which can be used to test specific predictions of theoretical models for neutrino mixing,
and which allow to highlight the impact of recent appearance data.

Figure~4 shows the joint contours at 1, 2 and $3\sigma$ ($\Delta \chi^2=1$, 4 and 9) for each
possible couple of $\sin^2\theta_{ij}$ parameters,  
in the analysis with old reactor fluxes. Including new fluxes, the best fits and the associated $N\sigma$ contours
are all translated by small amounts ($<1\sigma$) indicated by arrows. As a result of the dominance of T2K data in the
$\theta_{13}$ fit, 
the correlation in the $(\sin^2\theta_{12},\,\sin^2\theta_{13})$ plane induced by $\delta m^2$-sensitive
data \cite{HINT,Ve09} is no longer apparent. Conversely, there is a weak anticorrelation 
in the $(\sin^2\theta_{23},\,\sin^2\theta_{13})$ plane for relatively high $\theta_{13}$, due to the  fact that the
long-baseline $\nu_\mu\to\nu_e$ appearance probability is dominated by the product
$|U_{\mu3}U_{e3}|^2\propto \sin^2\theta_{23}\sin^2 2\theta_{13}$. 
{\color{black} If we had we hypothetically obtained a best fit value $\sin^2\theta_{23}\simeq 0.5$ 
instead of $0.42$, the best-fit value of $\sin^2\theta_{13}$ would have been presumably slightly lower (by 
about $-0.002$, according to our educated guess). }

\newpage

%\vspace*{-14cm}
%%%%%%%%%%%%%%%%%%%%%%%%%%%%%%%%%%%%%%%%%%%%%%%%%%%%%%%%%%%%%%%%%%%%%
\begin{figure}[t]
\vspace*{-1.7cm}
\includegraphics[width=0.64\textwidth]{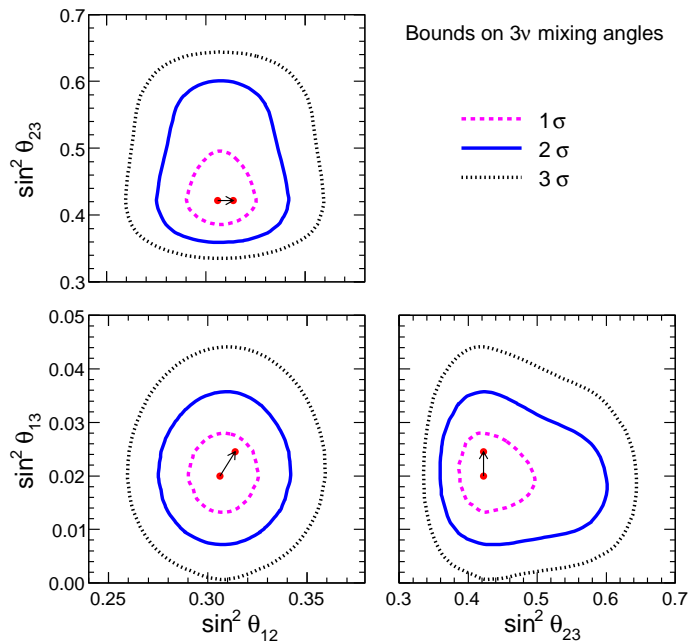}
\vspace*{-1.2cm}
\caption{\label{fig_4} 
Global $3\nu$ analysis: Joint contours at 1, 2 and $3\sigma$ ($\Delta \chi^2=1$, 4 and 9) for couples of
$\sin^2\theta_{ij}$ parameters, assuming old reactor neutrino fluxes. For new reactor fluxes,
the best fits (and, to a large extent, also the contours) are shifted as indicated by the arrows. }
\end{figure}
%%%%%%%%%%%%%%%%%%%%%%%%%%%%%%%%%%%%%%%%%%%%%%%%%%%%%%%%%%%%%%%%%%%%%%%

All the results reported in Figs.~1--4 and in Table~I are marginalized over the four discrete cases in 
Eq.~(\ref{disc}). For completeness, we also show in Fig.~5 the breakdown of the $\theta_{13}$ bounds over 
these four options (for the case of old reactor fluxes only), in terms of standard deviation
ranges for the parameter $\cos\delta\sin\theta_{13}=\pm\sin\theta_{13}$ in both normal and inverted
hierarchy, for the relevant data set ATM+LBL+CHOOZ
(while $\delta m^2$-sensitive data are $\delta$-independent). 
We do not find any significant difference in the position and likelihood of the best fits 
points for normal and inverted hierarchy. However, we find, as in our previous analysis \cite{Fo06},
a preference for the CP-conserving case $\cos\delta=-1$ versus $+1$. In \cite{Fo06}
we argued that the weak preference for both 
$\sin^2\theta_{13}>0$ and $\cos\delta <0$ were tied by the 
interference term driven by $\delta m^2$ and $\theta_{13}$ \cite{Pere} 
in the atmospheric neutrino oscillation probability.
The  preference for 
$\cos\delta\sin\theta_{13}<0$ in Fig.~5 appears to be more pronounced ($\sim 2\sigma$) than in \cite{Fo06}, 
presumably because it correlates with the (now more robust) preference for $\sin^2\theta_{13}>0$. It is
intriguing to notice that a weak preference for $\cos\delta\sin\theta_{13}<0$ has also been
found in the preliminary atmospheric $3\nu$ analysis from the SK collaboration \cite{Nu10}. 
Time will tell if this is another hint or mere fluctuation.

%%%%%%%%%%%%%%%%%%%%%%%%%%%%%%%%%%%%%%%%%%%%%%%%%%%%%%%%%%%%%%%%%%%%%
\begin{figure}[b]
\vspace*{-2.8cm}
\includegraphics[width=0.45\textwidth]{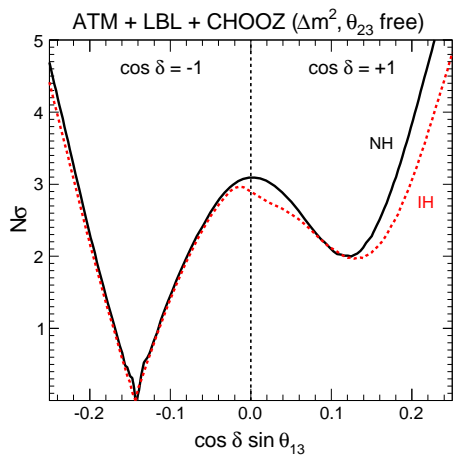}
\vspace*{-1.6cm}
\caption{\label{fig_5} 
Standard deviations from the best fit in terms of the variable
$\cos\delta\sin\theta_{13}$ for the two CP parities ($\cos\delta=\pm1)$ and for both
normal hierarchy (NH) and inverted hierarchy (IH), using the ATM+LBL+CHOOZ data set. 
The curves refer to the analysis with old
reactor neutrino fluxes; similar results (not shown) are obtained for new reactor fluxes. }
\end{figure}
%%%%%%%%%%%%%%%%%%%%%%%%%%%%%%%%%%%%%%%%%%%%%%%%%%%%%%%%%%%%%%%%%%%%%%%

\newpage

{\color{black} As a final remark, we note that the evidence in favor of $\theta_{13}>0$, found at a level
$>3\sigma$ in this paper, could only be strengthened by taking the CP phase $\delta$ as a free, continuous parameter.
Indeed, for $\theta_{13}=0$, the value of $\delta$ is physically irrelevant, and thus the value of 
$\chi^2_0=\chi^2|_{\theta_{13}=0}$ remains unchanged. For $\theta_{13}>0$ and unconstrained $\delta$, 
the absolute minimum of $\chi^2$ can only decrease 
(as compared to our constrained cases, $\cos\delta=\pm1$), leading to an increase
of $\Delta\chi^2=\chi^2_0-\chi^2_{\min}$, and thus to
an even higher statistical evidence in favor of $\theta_{13}>0$.  Of course, for free $\delta$
the best fit of $\sin^2\theta_{13}$ might be reached at values slightly different than 
those discussed herein (0.021 or 0.025, depending on reactor systematics).}
 
%\vspace*{-3mm}
\section{Discussion and conclusions} 

The hints of $\theta_{13}>0$ that we pointed out in \cite{NOVE,HINT} are consistent with recent T2K \cite{TtoK} and MINOS
\cite{MIN2} data and, in combination, provide an evidence for $\theta_{13}>0$ at the level of $>3\sigma$.
Such evidence, and the preference for values in the range 
$\sin^2\theta_{13}\simeq 0.01$--0.04 at $2\sigma$ (see Table~I) may
have far-reaching consequences in neutrino physics. First of all, it is crucial to experimentally test
these findings, not only with further long-baseline appearance data at accelerators, but also with
with short-baseline disappearance searches at reactors \cite{Me10}. If confirmed, the evidence 
for $\sin^2\theta_{13}\sim\mathrm{few}\,\%$ would open the door to CP violation searches in the neutrino sector,
with profound implications for our understanding of the matter-antimatter asymmetry in the universe.  
At the same time, matter effects on $(\pm\Delta m^2,\,\theta_{13})$-driven oscillations appear now
as a much more promising tool, from both the experimental and theoretical viewpoint, to 
derive indications about the hierarchy from neutrino propagation in the Earth or in supernovae.
Concerning nonoscillation searches, the evidence for $\theta_{13}>0$ provides small but
``guaranteed'' contributions of the third neutrino mass $m_3$ to both single beta and double beta decay
searches, which need to be accounted for in detailed analyses. 
From a more theoretical viewpoint, relatively large values for
$\theta_{13}$ will certainly trigger new ideas for model building.

%\vspace*{-3mm}
%%%%%%%%%%%%%%%%%%%%%%%%%%%%%%%%%%%%%%%%%%%%%%%%%%%%%%%%%%%%%%%%%%%%%%%%%
\acknowledgments
G.L.F., E.L., A.M., and A.M.R.\ acknowledge 
support by the Italian MIUR and INFN through the ``Astroparticle Physics'' 
research project. 
The work of A.P.\ is supported by the DFG Cluster of Excellence on the ``Origin and Structure of the Universe.''
E.L.\ thanks M.\ Goodman and M.\ Mezzetto for useful communications.

%%%%%%%%%%%%%%%%%%%%%%%%%%%%%%%%%%%%%%%%%%%%%%%%%%%%%%%%%%%%%%%%%%

\end{document}